# Super-Cerenkov Radiation as New Exotic Decay in Refractive Media

D. B. Ion[1,2)] and M. L. D. Ion[3)]

[1)] National Institute for Physics and Nuclear Engineering Horia Hulubei, IFIN-HH,
Bucharest, P.O.Box MG-6, Magurele Romania

[2)] Academy of Romanian Scientist (A.O.S.R.)

[3)] Faculty of Physics, Bucharest University, Bucharest, Romania

## Abstract

Generalized Super-Čerenkov Radiations (SČR), as well as their SČR-signatures are investigated. Two general SČR-coherence conditions are found as two natural extremes of the same spontaneous particles decays in (dielectric, nuclear or hadronic) media. The main results on the experimental test of the super-coherence conditions, obtained by using the experimental data from BNL, are presented. The interpretation of the observed anomalous Čerenkov rings as experimental evidence for the HE-component of the SČR is discussed.

**Key words**: Čerenkov radiation, Super-Čerenkov effect, Anomalous Čerenkov rings, Nuclear pionic Čerenkov-like radiation (NPIČR), particle refractive index.

## 1. Cerenkov Radiation (CR)

The classical theory of the radiation emitted by charged particles moving with superluminal velocities were traced back to Heaviside [1]. In fact, Heaviside considered the Čerenkov radiation [2] in a nondispersive medium. He considered this topic many times over the next 20 years, deriving most of the formalism of what is now called Čerenkov radiation and which is applied in the particle detectors technics (e.g., RICH-detectors). So, doing justice (see the papers of Kaiser and Jelley in Nature) to Heaviside [1] De Coudres [3] and Somerfeld [4], we must recall that the classical theory of the CR phenomenon in a dispersive medium was first formulated by Frank and Tamm in 1937 [5]. This theory explained all the main features of the radiation observed experimentally by Čerenkov [2] (see Fig. 1). In fact, from experimental point of view, the electromagnetic Čerenkov radiation was first observed in the early 1900's by the experiments developed by Marie and Pierre Curie when studying radioactivity emission. In essence they observed that phenomenon consists from *the very faint emission of a bluish-white light from transparent substances in the neighborhood of strong radioactive source.* But the first deliberate attempt to understand the origin of this light was made by Mallet [6] in 1926-1929. He observed that this *light emitted by a variety of transparent bodies placed close to a radioactive source always had the same bluish-white quality, and that the spectrum was continuous, not possessing the line or band structure characteristic of fluorescence.* Mallet found such an emission, he could not offer an explanation for the nature of this phenomenon, but was the first to find that spectrum was continuous and extended to 3700 Å. He was the first to appreciate the universality of this effect but no attempted to study the polarization of this new kind of radiation.

Only the exhaustive experimental work, carried out between years 1934-1937 by P. A. Čerenkov [2], characterized completely this kind of radiation. These experimental data are fully consistent with the classical electromagnetic theory developed by Heaviside in 1888 [1] and Frank and Tamm [5] (see Fig.1). In essence, it was revealed by the Heaviside, Čerenkov, Tamm and Frank that a charged particle moving in a transparent medium with an refractive index, $n_\gamma$, and having a speed $v_1$ greater than phase velocity of light $v_{\gamma ph} = n_\gamma^{-1}$ will emit electromagnetic radiation, called Čerenkov radiation (ČR), at an polar emission angle $\theta_C$ relative to the direction of motion given by the relation (we adopted the system of units $\hbar = c = 1$ ):

$$\cos\theta_C = \frac{v_{\gamma ph}(\omega)}{v_x} \leq 1 \qquad (1)$$

The remarkable properties of the Čerenkov radiation find wide applications in practice especially in high energy physics where it is extensively used in experiments for counting and identifying relativistic particles [via Ring Imaging Čerenkov (RICH)-Detectors,e.g. see: T. Ypsilantis and J. Seguinot, *Nucl. Instrum. Methods A* 433, 1 (1999)] in the fields of elementary particles, nuclear physics and astrophysics. A quantum theoretical approach of the Čerenkov effect by Ginsburg [7] resulted in only minor modification to the classical theory [see also the books [8-9]]. Some interesting discussions about the predictions and experimental discovery of the Čerenkov radiation can be found in the papers of Kaiser [10], Jelley [11], Tyapkin [12] and Govorkov [13].

Now, the Čerenkov radiation (ČR) is the subject of many theoretical and experimental studies related to the extension to the nuclear and hadronic media as well as to other coherent particle emission via Čerenkov-like mechanisms [14-56]. The generalized Čerenkov-like effects based on four fundamental interactions has been investigated and classified recently in [44]. In particular, this classification includes the nuclear (mesonic, $\gamma$, weak boson)-Čerenkov-like radiations as well as the high energy component of the coherent particle emission via (baryonic, leptonic, fermionic) Čerenkov-like effects..

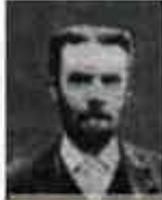
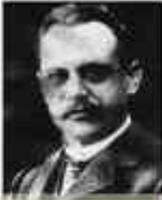
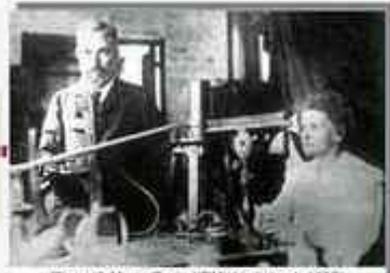
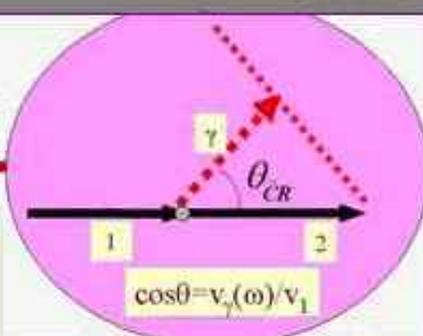

Fig.1 A brief history of Cerenkov radiation

However, by recent experimental observations of the subthreshold and anomalous Cerenkov radiations (CR) (see Fig. 2a,b) as well as multi-ring phenomena it was clarified that some fundamental aspects of the CR can be considered as being still open and that more theoretical and experimental investigations on the CR are needed.

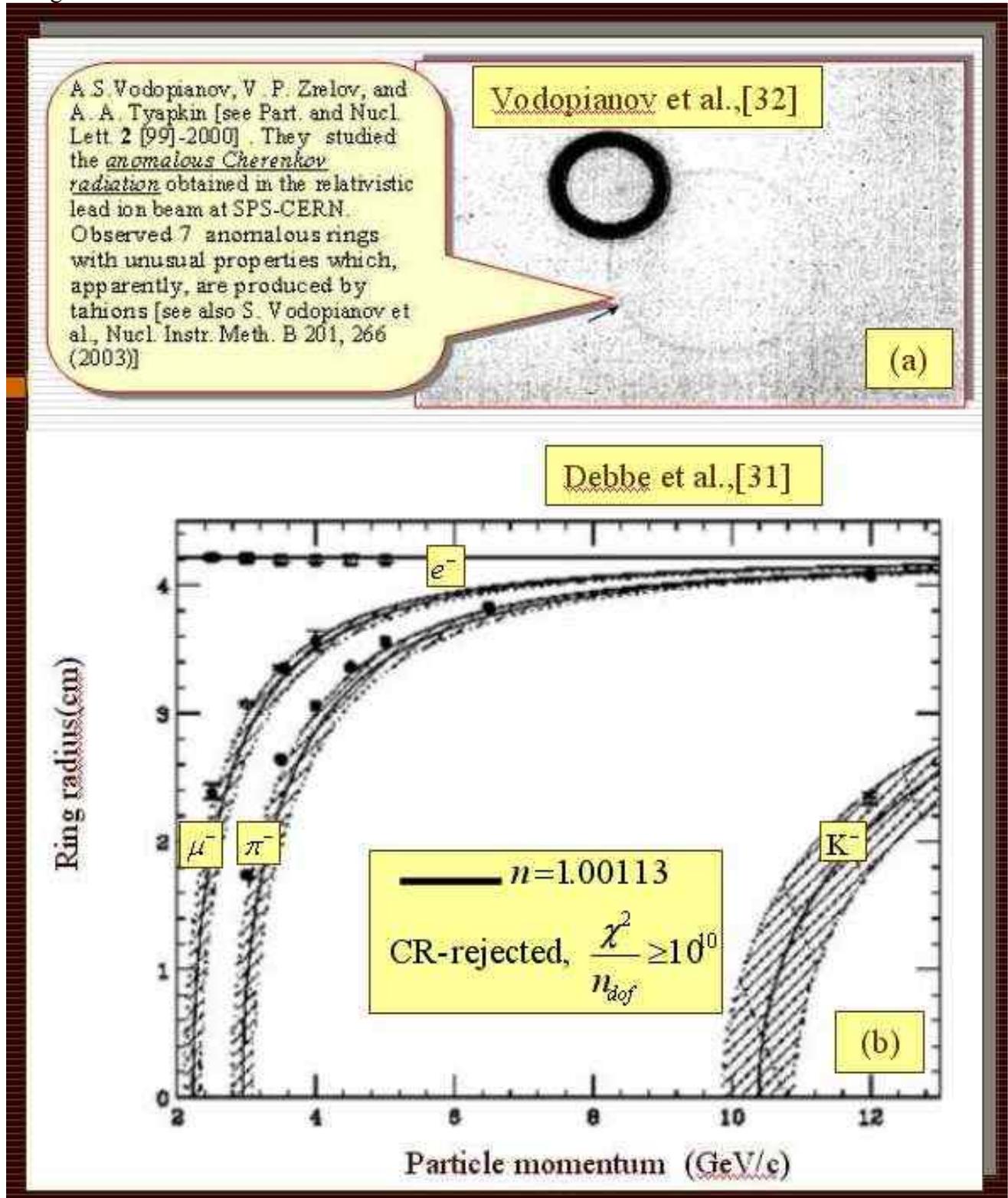

Fig.2: (a) An example of anomalous Cerenkov rig observed by Vodopianov et al., [32] CERN SPS accelerator, (b) Experimental observation [31] at BNL of significant discrepancy between the experimental Cerenkov rings radii and absolute theoretical CR-predictions (solid line) are compared and the CR-hypothesis rejected.

## 2. Super-Cerenkov Radiation

As we already mentioned, the new results about subthreshold CR, anomalous CR-rings (see Fig. 2) and concentric multi-CR rings (see Fig.3) stimulated new investigations about the origin of CR-coherence condition (see Fig. 1). Then, theoretical investigations using the CR correct kinematics lead us the discovery that CR is in fact only the low energy component of a more general phenomenon called by us the Super-Cerenkov radiation (SCR) characterized by the Super-Cerenkov (SCR)-decay condition presented in Figs.3-4.

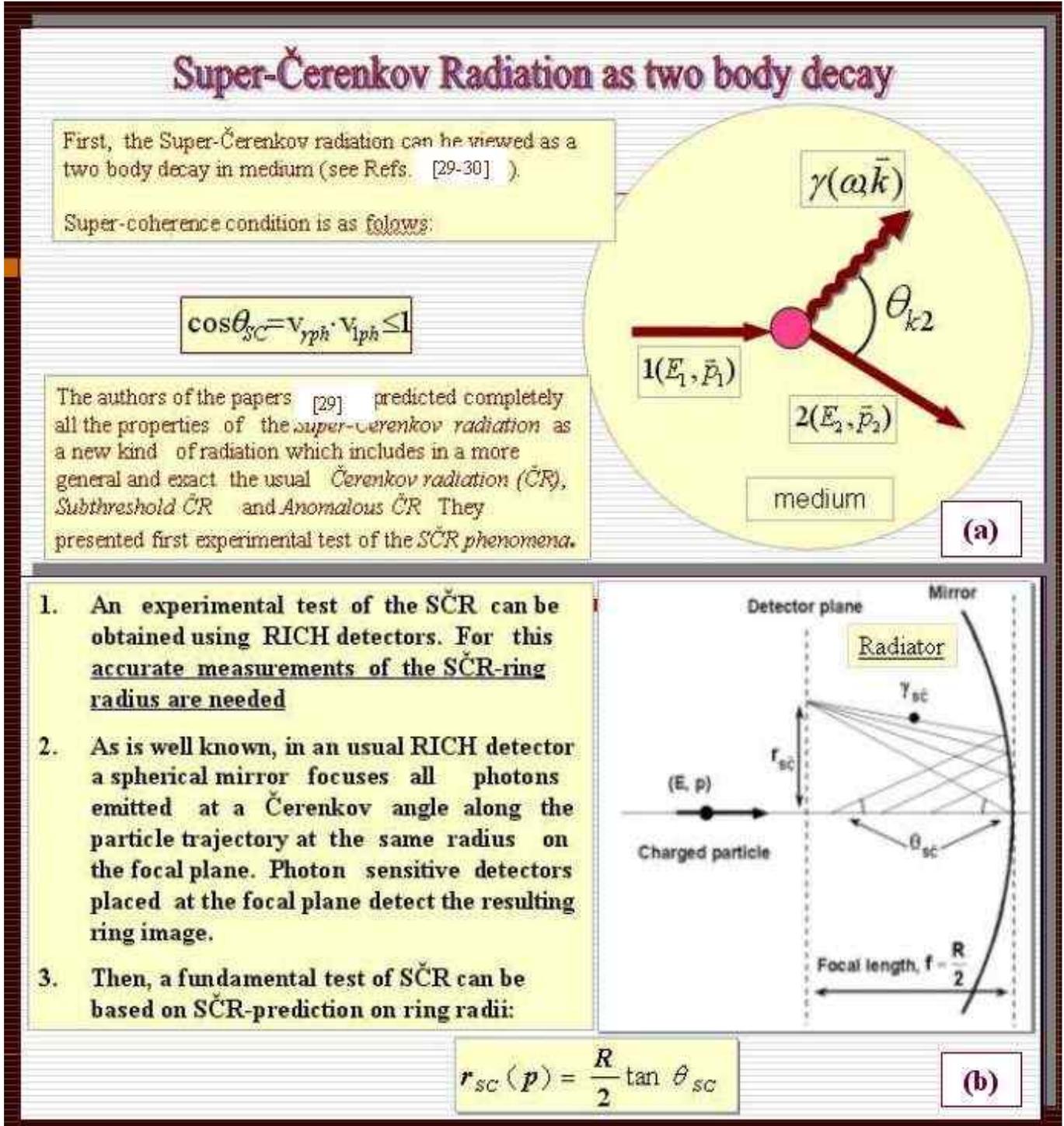

Fig.3: (a) Schematic description of Super-Cerenkov radiation. (b) Schematic description of a RICH detector.

## SCR-coherence conditions

From energy-momentum conservation laws $E_1 = \omega + E_2$, $\mathbf{p}_1 = \mathbf{k} + \mathbf{p}_2$, we obtain

- SČR-angle theta(1k)

$$\cos \theta_{1k} = v_{1ph} v_{\gamma ph} + \frac{D_2 - D_1 - D_\gamma}{2 p_1 k}$$

- SČR-angle theta(12) and theta(2k)

$$\cos \theta_{12} = v_{1ph} v_{2ph} + \frac{D_\gamma - D_1 - D_2}{2 p_1 p_2},$$

$$\cos \theta_{2k} = v_{2ph} v_{\gamma ph} + \frac{D_1 - D_2 - D_\gamma}{2 p_2 k},$$

where $D_x \equiv E_x^2 - p_x^2$

- SČR semiclassical decay condition for $\theta_{1k}$:

$$\cos \theta_{1k} = v_{1ph} v_{\gamma ph} \leq 1$$

- SČR semiclassical decay condition for $\theta_{12}$ and $\theta_{2k}$:

$$\cos \theta_{12} \approx v_{1ph} v_{2ph} \leq 1$$

$$\cos \theta_{2k} \approx v_{2ph} v_{\gamma ph} \leq 1$$

Fig.4: A short proof of the Super-Cerenkov coherence condition (2).

As we see, from Fig. 4, the SCR-condition (2) is obtained in a natural way from the energy-momentum conservation law when the influence of medium on the propagation properties of the charged particle is also taken into account.

$$\cos \theta_{SC} = v_{xph}(E_x) \cdot v_{\gamma ph}(\omega_\gamma) \leq 1 \qquad (2)$$

where $v_{xph}(E_{xi})$ and $v_{xph}(E_{xf})$ are phase velocities of the charged particle x in the initial and final states, respectively.

### 2.1 Summary of SCR-signatures

Indeed, the *super-Čerenkov relation* (2) can be easy proved by using the energy-momentum conservation law for the "decay" $1 \rightarrow 2 + \gamma$ (see Fig..4) to obtain the SČR-angles as they are given in Fig. 5. The signature of the SČR-effects are schematically described in Fig.10. Two components, corresponding to [low gamma energy and high gamma energy]-emissions, with the orthogonal gamma polarizations, are clearly evidentiated

(a) Low $\gamma-energy$ sector or Čerenkov radiation sector (see Fig.5, upper part):

$$\cos\theta_{SC}=\cos\theta_{1\gamma}=v_{\gamma ph}.v_{1ph}\Rightarrow \cos\theta_{1\gamma}=\frac{v_{\gamma ph}(\omega)}{v_1\,\mathrm{Re}\,n_1}\leq 1 \qquad (3)$$

(b) High $\gamma-energy$ or "*Source*" Čerenkov-like spontaneous bremsstrahlung sector (see Fig. 5, lower part):

$$\cos\theta_{SC}=\cos\theta_{12}=v_{1ph}.v_{2ph}\Rightarrow \cos\theta_{12}=\frac{v_{2ph}(E_2)}{v_1\,\mathrm{Re}\,n_1}\leq 1 \qquad (4)$$

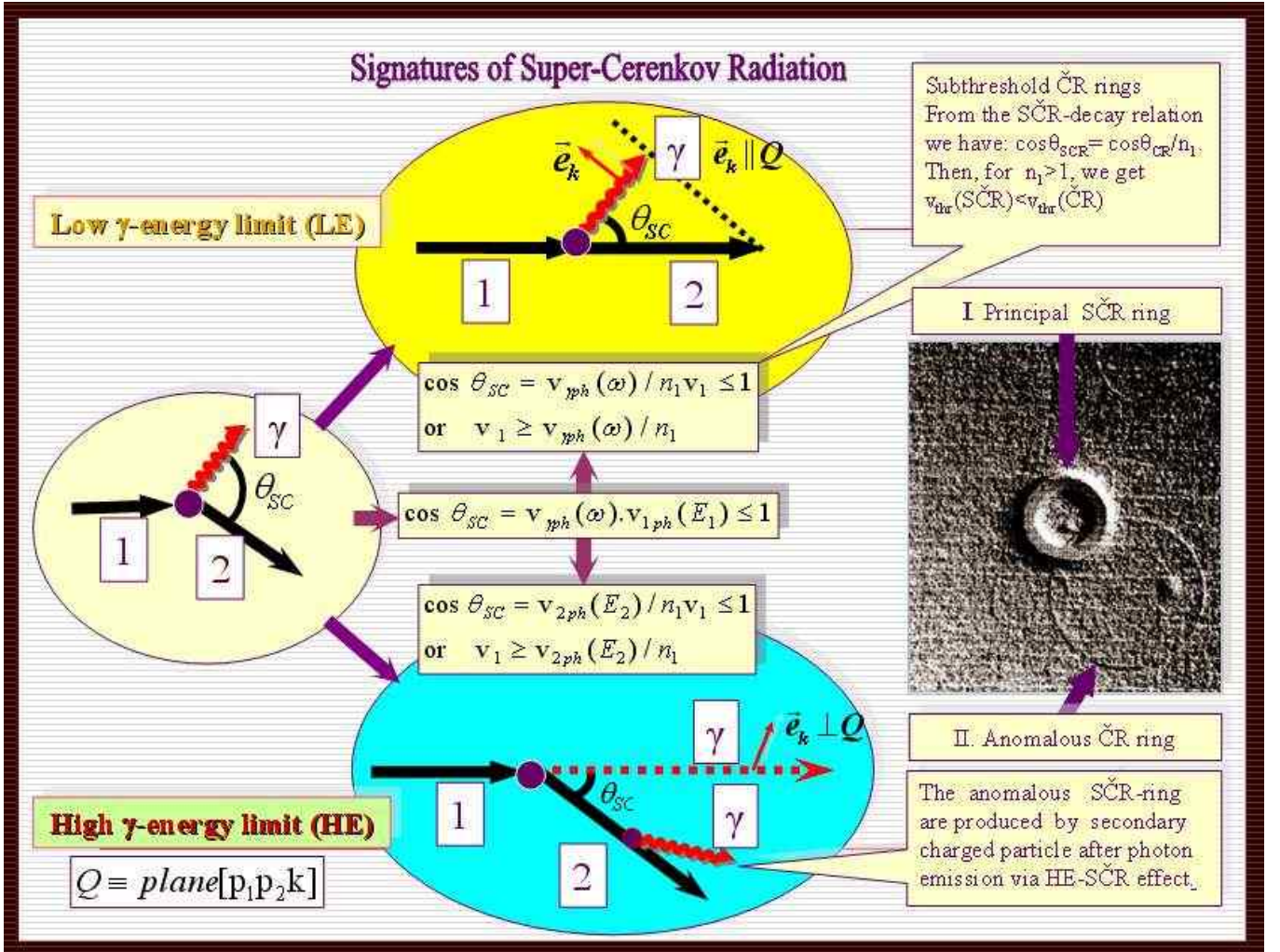

Fig.5: The principal signatures of the Super-Cerenkov Radiation (SCR), as they are obtained from the quantum SCR-theory, are summarized. The principal SCR-beam ring and anomalous SCR-anomalous ring [32] are also presented (see Section 2.4)

The quantum theory of SČR is similar to the quantum theory [9] of the usual Čerenkov Radiation and here we present only some final results for the case of $\gamma$-decays of spin ½-particles in (dielectric, nuclear, hadronic)-transparent nondispersive media. So, just as in the quantum ČR-theory the same interaction Hamiltonian $H_{fi}$ with some modifications of source fields in medium can also describe the coherent γ-emission in all sectors (LE and HE). Then it is easy to see that the intensity of Super-Cerenkov radiation can be written as in Fig.6, where the spin factor S, for a two body electromagnetic "SCR-decay" of a spin ½-particle in medium is given by:

$$S\equiv\frac{(E_1+M)(E_2+M)}{4E_1E_2}\left[\frac{p_1^2}{E_1+M}+\frac{p_2^2}{E_2+M}+2\frac{(\vec{e}_k\cdot\vec{p}_1)(\vec{e}_k\cdot\vec{p}_2)-(\vec{e}_k\times\vec{p}_1)(\vec{e}_k\times\vec{p}_2)}{(E_1+M)(E_2+M)}\right] \qquad (6)$$

Now, one can see that $\Theta(1-\cos\theta_{SC})$-Heaviside step function is 1 in two (or many) physical regions defined by the constraints: $\cos\theta_{2\gamma} = v_{2ph}(E_2) v_{\gamma ph}(\omega) \leq 1$. The spin factor S in the above formulas is defined just as in the usual ČR-quantum theory but with the particle's momentum $p_i$, i= 1,2, considered in medium. The vector $\mathbf{e}_k$ is the photon polarization for a given photon momentum $\mathbf{k}$. For a given $\mathbf{k}$ we choose two orthogonal photon spin polarization directions, corresponding to a polarization vector perpendicular and parallel to the SČR-decay plane Q. Then, we obtain the result given in Fig.6. for low γ-energy (LE) and high γ-energy limits, respectively.

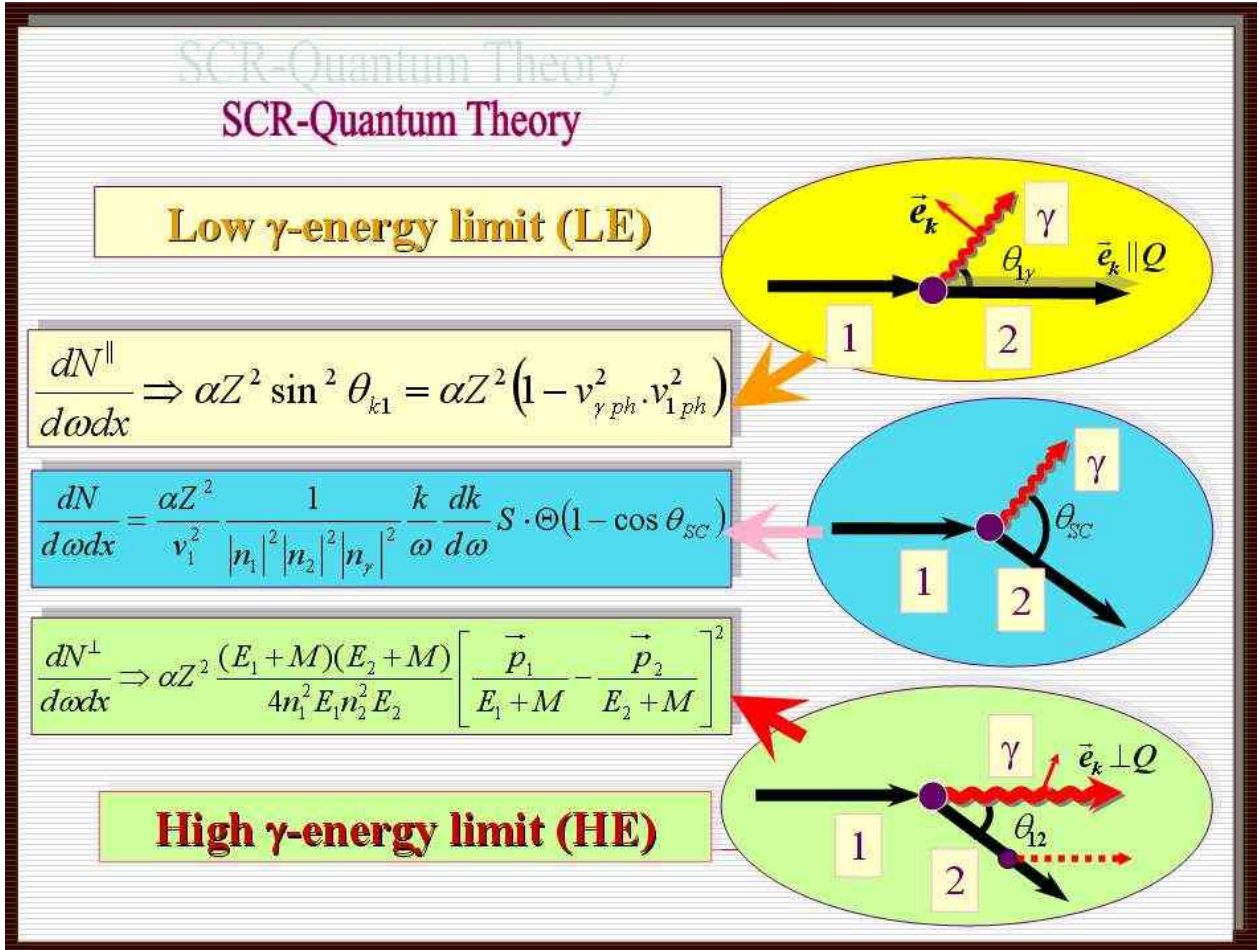

Fig. 6 The main predictions of SCR-Quantum Theory for the exotic electromagnetic decays of spin ½-particles in (dielectric, nuclear, hadronic)-media.

### 2.2. Experimental Tests of Super-Čerenkov decay condition (2)

Čerenkov radiation is extensively used in experiments for counting and identifying relativistic particles in the fields of elementary particles, nuclear physics and astrophysics. A spherical mirror focuses all photons emitted at Čerenkov angle along the particle trajectory at the same radius on the focal plane as is shown in Fig.8. Photon sensitive detectors placed at the focal plane detect the resulting ring images in a Ring Imaging Čerenkov (RICH) detector. So, RIČH-counters are used for identifying and tracking charged particles. Čerenkov rings formed on a focal surface of the RIČH provide information about the velocity and the direction of a charged particle passing the radiator. The particle's velocity is related to the Čerenkov angle $\theta_C$ [or more exactly to the Super-Čerenkov angle $\theta_{SC}$ ] by the relations (1) [or (2)], respectively. Hence, these angles are determined by measuring the radii of the rings detected with the RIČH. In Ref. [31] a $C_4 F_{10} Ar(75:25)$ filled RIČH-counter read out (by a 100-channel photomultipiler of $10\times 10$ cm² active area) was used for measurement in beams of the Čerenkov ring radii for electrons, muons, pions and kaons.

Therefore, the problem of the experimental test of Super-Čerenkov coherence condition is of great interest not only for the fundamental physics but also for practical applications to the particle detection. Recently such a test was performed by us [29,30] by using the experimental data of Debbe et al. [31] from the Ring Imaging Čerenkov (RIČH) detectors obtained at BNL. The results are presented in Fig.7.

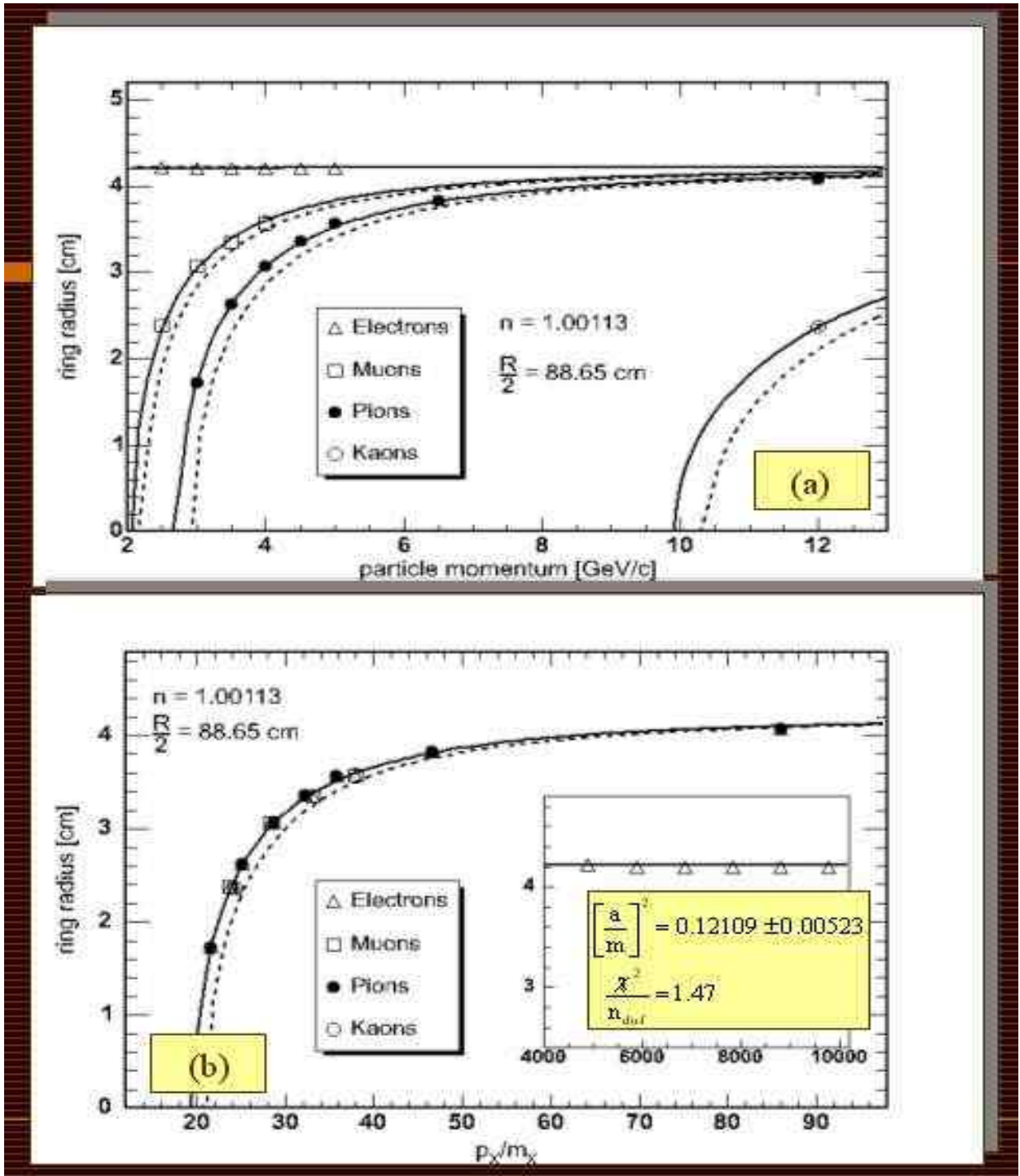

Fig.7: Experimental Čerenkov ring radii of the particles e, µ, π, K, obtained by Debbe et al. [31] with RICH detector, are compared with the theoretical [29] *Super-Čerenkov predictions* (solid curves ), and also with the *Čerenkov predictions* (dashed curves,)

**2.3. Anomalous Čerenkov Rings as important Signature of Super-Čerenkov Radiation**

The Čerenkov radiation caused by relativistic lead ions was studied at SPS CERN [32]. A beam of $^{208}Pb^{82+}$ ions with the energy of 157.7 A GeV was going along the axis of the Čerenkov detector. The Čerenkov light emitted in the radiator (its length along the optical axis is 405 mm) got into the objective of a photocamera after its refection in the mirror inclined under 45 degree angle relative to the axis of the radiator (see the original paper [32] for details). A bright narrow ring of the Čerenkov radiation seen on the picture in Fig.8 is caused by relativistic lead ions. Besides, in this picture we have found hardly noticeable narrow Čerenkov radiation rings of the particles flying out under small angles to the direction of the beam. The calculation of the velocity of these particles has shown that it corresponds to those of particles moving faster than the light velocity in the vacuum.

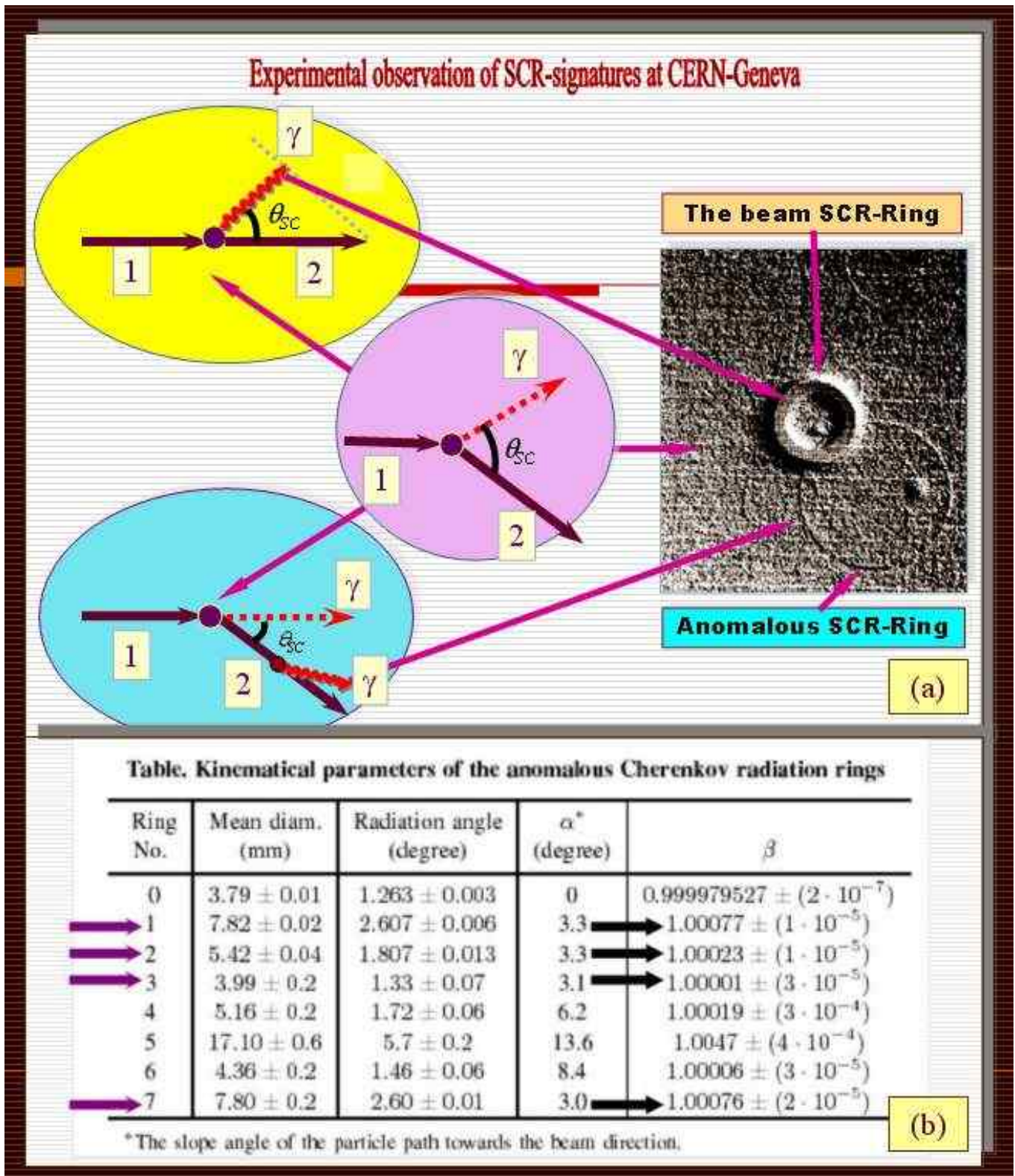

Fig.8: Experimental observation [32] of seven anomalous SČR-rings at CERN-SPS accelerator from Geneva. (a) An example of anomalous Čerenkov ring (ring No. 7) compared with the usual beam ČR-ring .

The conclusion of Vodopianov et al [32] is that the large Čerenkov radiation ring, shown with Ring number 7 in Fig. 8b, corresponds to a tachion velocity approximately equal to $\beta \approx 1.0008$. The ring diameter of its radiation is approximately two times larger than the ring diameter of the proton radiation at the velocity of motion $\beta \to 1$. Totally, seven rings of the anomalous Čerenkov radiation have been found in the three photos [32]. The rings were analyzed by Vodopianov et al. [32] using the standard approximation expression:

$$\cos\theta = 1/n_\gamma v \tag{7}$$

instead with the true SCR complete formula

$$\cos\theta = \frac{1}{n_{Pb^+}(E) n_\gamma(\omega) v} \tag{8}$$

where $n_{Pb^+}(E)$ is the refractive index of the lead ions $^{207}_{82}Pb^+$ in medium. Therefore, the authors of Ref. [58] obtained in their table the values for $\beta n_{Pb^+}$ instead the values of $\beta$. Therefore, the observation of the anomalous Čerenkov ring can be interpreted as one of the most important signature of the HE-component of Super-Čerenkov radiation (see Figs. 5 and 8) produced by lead ions in the radiator medium.

### 2.4. Astrophysical implication of two-components of Super-Cerenkov radiation

Detailed astrophysical implication of the gamma SČR-predictions is also investigated in our project ID-52-283/2007. Then, we known in advance that the astonishingly similar character of the pulse profile (see that of the Crab Pulsar) in very different spectral regions of Pulsars can be well accounted for by the two-component Super-Čerenkov radiations. So, different strongly correlated (high and low) SČR-bands can be emitted by the same beam of charged particles simultaneously in the same time and space region of a pulsar. Therefore, the same "pulse profile" will be obtained as a simple consequence of the simultaneous emission of all SČR- Čerenkov-like bands. We also investigated the possibility of SČR-emission from Pulsars in the paper [25]. So, very high energy gamma radiation (VHEGR) ( $E_\gamma \geq 100\,GeV$ ) and ultrahigh energy gamma radiations (UHEGR) ( $E_\gamma \geq 100\,TeV$ ) are produced by ultra-relativistic charged particles (e.g. electrons, muons, protons, light nuclei, etc.) during their interaction with ambient medium (atoms, nuclei, etc.). Then, we obtained that the SCR predictions are satisfied experimentally to a surprising accuracy by the data from some important VHEGR/UHEGR sources such as Crab and Vela Pulsars, Cygnus X-3 and Vela X-1 binary systems, etc. The astonishingly similar character of the pulse profile is also well explained by the Super-Čerenkov two components mechanism in nuclear media since different spectral Čerenkov bands can be simultaneously produced by the same beam of accelerated particles in the same space region of the pulsar.

### 3. Mesonic Čerenkov-like effects in hadronic and nuclear media.

The idea that meson production in nuclear interactions may be described as a process similar to the Čerenkov radiation has considered [14-25] by Wada (1949), Ivanenko (1949) Blohintev si Indenbom (1950), Čzyz, Ericson, Glashow (1959), Smrz (1962) si D. B. Ion.(1969-1970). For the many detailed results on mesonic Cerenkov-like effect see Refs.[15,16,20,28,33] and the results from Fig.9 .
The classical variant of the theory of the mesonic Čerenkov-like radiation in hadronic media [21] was applied to the study of single meson production in the hadron-hadron interactions at high energy. This variant is based on on the usual assumption that hadrons are composed from a central core (in which the hadron mass is concentrated surrounded by a large and more difuse mesonic cloud. (hadronic medium). Then it was shown [15,16,20,28,33] that a *hadronic mesonic Čerenkov-like radiation* (HMČR) mechanism (see [15,16,20,28,33]), with an mesonic refractive index in hadronic medium given by pole approximation, is able to describe with high accuracy the integrated cross section of the single meson production in the hadron-hadron interactions.

To illustrate these important results in Fig. 2b we presented the measured integrated cross section for the process: $pp \to pp\pi^0$ compared 15,16,20,28,33] with the prediction of mesonic Čerenkov-like radiation (HMČR-mechanism). This result was very encouraging for the extension of the Čerenkov-pions analysis (HMČR-variant) to all processes of single meson production in hadron-hadron interaction. The results of such analyses are presented in our papers [15,16,20,28,33]]. Collecting $\chi^2/dof$ for all 139 reactions fitted with the HMCR approach [15,16,20,28,33] we obtained the surprising results presented in Fig. 9b.

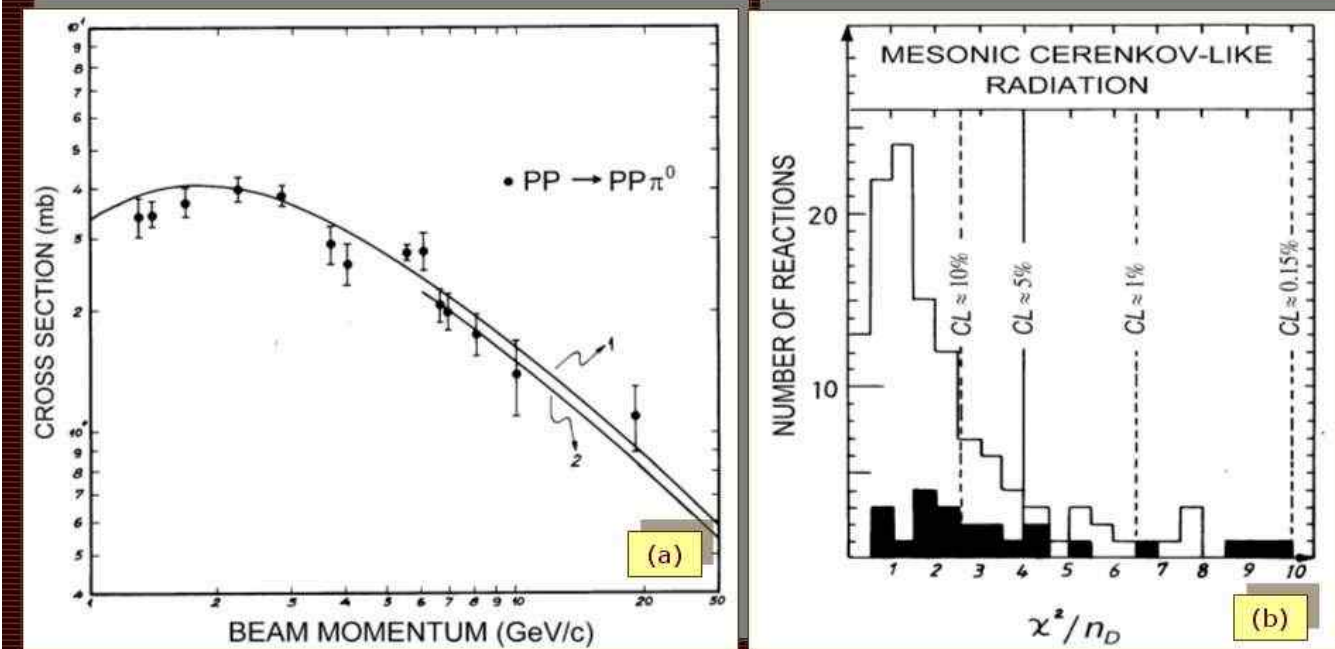

**Fig. 9:** Experimental evidences for Cerenkov mesons in hadronic media. (a) measured integrated cross section for the process $pp \to pp\,\pi^0$ compared [15] with the prediction of mesonic Cherenkov-like radiation in hadronic media (HMCR-mechanism). (b) The collection of $\chi^2/dof$ for all 139 reactions fitted with the HMCR approach [15,20,28]. The blacken region correspond to the reactions with single vector meson emission, while the white region correspond to the pseudoscar single meson production via HMCR-mecanism.

In 1990-19995, we have extended [19,21-25]] these ideas to the nuclear media where the pionic (NPIČR) and gamma Čerenkov radiation (NGČR) should be possible to be emitted from charged particles moving through nuclei with a velocity larger than the phase velocity of photons or/and pions in the nuclear media. The refractive indices of the gamma ($n_\gamma$), meson ($n_\pi$), nucleon ($n_N$), was calculated by using Foldy-Lax formula [34] and the experimental pion-nucleon cross sections combined with the dispersion relations predictions, the refractive index of pions in the nuclear media has been calculated [19,21-25]. Then, the detailed predictions for the spontaneous pion emission as nuclear pionic Čerenkov radiation (NPIČR) inside the nuclear medium are obtained and published in Refs. [19,21-25].

Moreover, it is important to note that in 1999, G.L.Gogiberidze, E.K. Sarkisyan and L.K. Gelovani [26] performed the first experimental test of the pionic Čerenkov-like effect (NPIČR) in Mg-Mg collisions at 4.3 GeV/c/nucleon by processing the pictures from 2m Streamer Chamber SKM-200. So, after processing a total of 14218 events, which were found to meet the centrality criterion, the following experimental results are obtained:
- The energy distribution of emitted pions in the central collisions have a significant peak (4.1 standard deviations over the inclusive background) (see Fig. 10).
- The value of the peak energy and its width are

$$E_m = [238 \pm 3(stat) \pm 8(syst)] MeV$$
$$\Gamma_\pi = [18 \pm 3(stat) \pm 5(syst)] MeV$$

So, they obtained a good agreement with the position and width of the first pionic Cerenkov-like band predicted by D.B.Ion and W. Stocker in ref. [25].

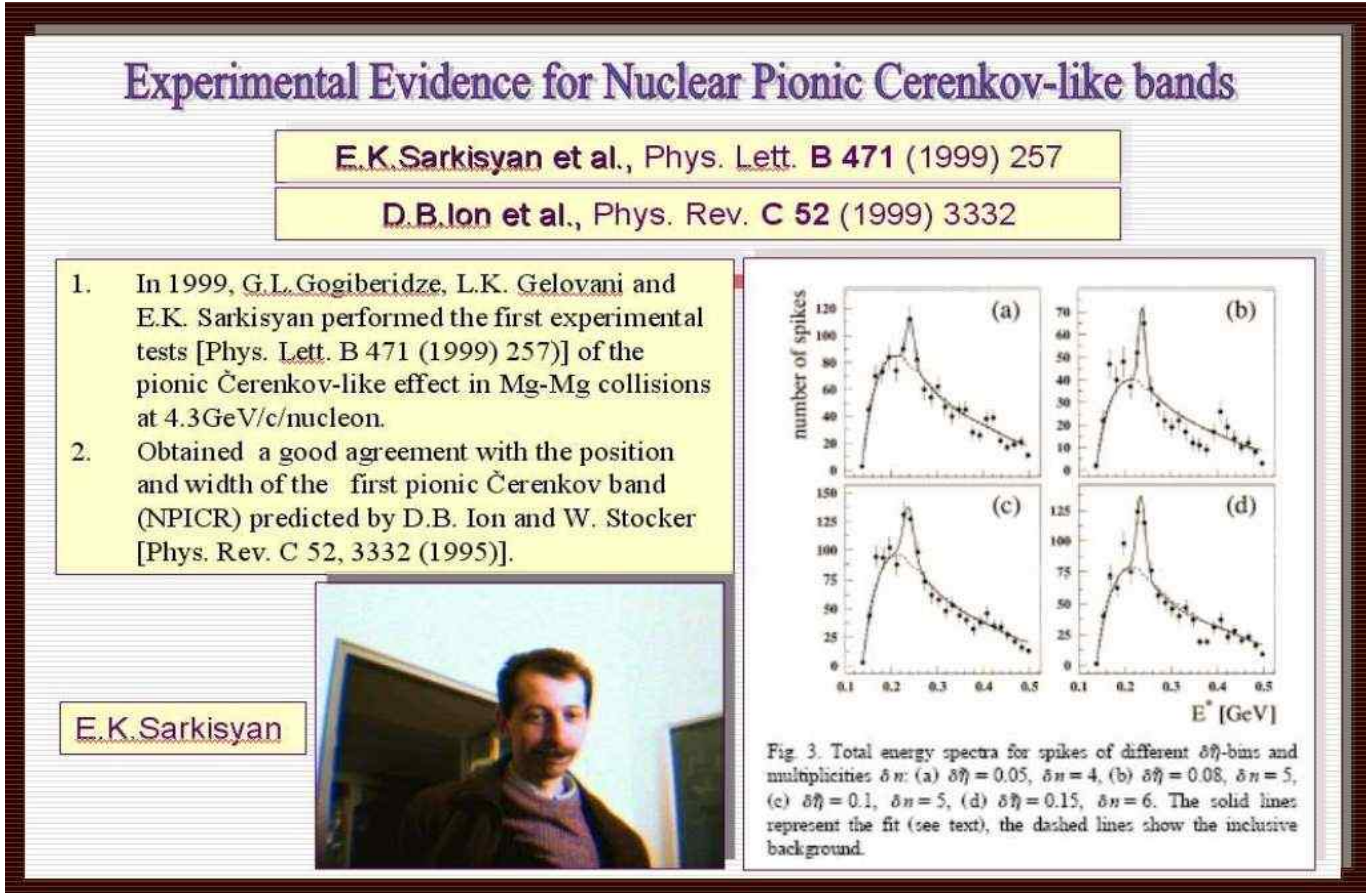

Fig. 10: Experimental evidences for NPIČR-pions in the first pionic-SČR band.predicted in Ref. [25].

## 4. Summary and Conclusions

By the recent experimental observations of the subthreshold and anomalous Čerenkov radiations (ČR) (see Figs. 2a,b) as well as multi-ring phenomena it was clarified that some fundamental aspects of the ČR can be considered as being still open and that more theoretical and experimental investigations are necessary. Then, theoretical investigations using the ČR correct kinematics lead us the discovery that ČR is in fact only the low energy component of a more general phenomenon called by us the Super-Čerenkov radiation (SCR) characterized by the Super-Čerenkov (SČR)-decay condition presented in Figs.3-4.

$$\cos\theta_{SC} = v_{\gamma ph} \cdot v_{xph} \leq 1$$

Our theoretical investigations shown that the SČR-phenomenon includes in an unified way :
(i) Gamma-Čerenkov radiation including subthreshold ČR (see Figs. 5-6,(*LE*)-*component*);
(ii) "Particle source" Čerenkov-like effect (see Fig.5-6, *HE-component*);
(iii) Anomalous Čerenkov radiation (secondary anomalous SČR-rings).

The experimental test of this SČR-coherence relation, near the usual Čerenkov threshold, was performed [29] by using the data of Debbe et al. [31] on Čerenkov ring radii of electrons, muons, pions and kaons in a RICH detector. Results on this experimental test of the super-coherence conditions are presented in Fig 7. These SČR-predictions are verified experimentally with high accuracy: χ2/dof=1.47.

The observation of the anomalous Čerenkov rings [32] was interpreted as one of the most important signature of the HE-component of Super-Čerenkov radiation (as shown in Fig.8) produced by lead ions in the radiator medium. The observed SCR-angles [ $\theta_{SC} \equiv \alpha \approx 3.1$ degree] between the lead pathway and the beam direction are consistent with the constant phase velocity of the final lead in the radiator medium, for all anomalous rings 1,2,3,7.

Finally, we remark that new and accurate experimental measurements of the Čerenkov ring radii, as well as for the anomalous HE-component of SČR are needed.

**Acknowledgments**

This research was supported by CNCSIS under contract ID-52-283/2007.